\documentclass[emulateapj]{aastex63}
\usepackage{float}
\usepackage{amsmath}
\usepackage{comment}
\usepackage{graphicx}
\usepackage[caption=false]{subfig}
\usepackage[flushleft]{threeparttable}
\usepackage{changepage}

\usepackage{kpfonts}
\begin{document}

\title{Precise Radial Velocities Using Line Bisectors}

\shorttitle{RVs Using Bisectors}
\shortauthors{Deming et al.}

\correspondingauthor{Drake Deming}
\email{ldeming@umd.edu}

\author[0000-0001-5727-4094]{Drake Deming}
\affiliation{Department of Astronomy, University of Maryland,
   College Park, MD 20742, USA}

\author[0000-0003-4450-0368]{Joe Llama}
\affiliation{Lowell Observatory, 1400 W Mars Hill Road, Flagstaff AZ 86001, USA}

\author[0000-0002-3263-2251]{Guangwei Fu}
\affiliation{Department of Physics and Astronomy, Johns Hopkins University, Baltimore, MD 21218 USA}

\begin{abstract}
  We study the properties of line bisectors in the spectrum of the Sun-as-a-star, as observed using the Integrated Sunlight Spectrometer (ISS) of the SOLIS project.  Our motivation is to determine whether changes in line shape, due to magnetic modulation of photospheric convection, can be separated from the 9 cm/sec Doppler reflex of the Earth's orbit.  Measuring bisectors of 21 lines over a full solar cycle, our results overwhelmingly indicate that solar magnetic activity modulates photospheric convection so as to reduce the asymmetries of line profiles in the spectrum of the Sun-as-a-star (having both C-shaped and reversed-C-shaped bisectors).  However, some lines are constant or have variations in shape that are too small to measure.

  We inject a 9 cm/sec radial velocity signal with a 1-year period into the ISS spectra.  Informed by a principal component analysis of the bisectors, we fit the most significant components to the bisectors of each line by linear regression, including a zero-point offset in velocity that is intended to capture the injected radial velocity signal.  Averaging over lines, we are able to recover that signal to solid statistical significance in the presence of much larger changes in the line shapes.  Although our work has limitations (that we discuss), we establish that changes in absorption line shapes do not in themselves prevent the detection of an Earth-like planet orbiting a Sun-like star using precise radial velocity techniques. 

\end{abstract}


\section{Introduction}

 A current frontier in exoplanetary science is the push to detect temperate planets of mass comparable to the Earth, orbiting stars similar to the Sun \citep{Fischer_2016, Morgan_2021}.  Given that the transit probability of such a planet is small ($<$1\%), and Sun-like stars are uncommon, detection of a nearby Earth-twin system using transits is unlikely.  Instead, the radial velocity (RV) technique is more appropriate, provided that the requisite sensitivity can be achieved.  
 
 The RV reflex of the Sun due to the orbit of the Earth is 9 cm/sec, approximately an order of magnitude below current detection capabilities.  The advent of laser frequency combs to stabilize RV spectrometers has greatly improved instrumental stability \citep{Halverson_2016}.  Intrinsic variations in the wavelengths of stellar lines, due primarily to magnetic activity are the limiting noise source for the RV detection of Earth-twins \citep{Crass_2021}. 
 
 In order to understand and mitigate effects of magnetic activity, and to provide proof-of-principle for extremely precise radial velocity (EPRV) techniques, the spectrum of disk-integrated sunlight (the 'Sun-as-a-star') plays a key role\footnote{Disk-resolved solar spectroscopy is also an important tool for understanding how magnetic activity affects RVs}.  The velocity stability of the Sun-as-a-star for the purpose of exoplanet detection has been of interest for decades \citep{Deming_1987, Wallace_1988, McMillan_1993, Deming_1994}, and interest has increased in recent years \citep{Lagrange_2010, Meunier_2015, Lanza_2016, Haywood_2016, Collier-Cameron_2019, Milbourne_2019, Haywood_2022}.   Several high precision stellar spectrometers now use solar-feed optics to monitor the integrated light solar spectrum on a daily basis \citep{Dumusque_2015, Strassmeier_2018, Dineva_2020, Lin_2022, Llama_2022}.

\subsection{Scope of This Work}

It is clearly possible to {\it distinguish} the effects of magnetic activity on spectral lines from center-of-mass motions \citep{Queloz_2001}.  Magnetic activity affects spectral line shapes \citep{Livingston_1983, Dravins_1990, Saar_1990, Cegla_2013, Cegla_2019}, whereas the Doppler reflex of a planet merely shifts spectral lines without affecting their shape (in velocity coordinates).  However, {\it cleanly separating} magnetic and center-of-mass shifts could be more problematic.  To that end, \citet{Collier-Cameron_2021} used solar integrated light spectroscopy from HARPS-North to demonstrate that magnetic and center-of-mass shifts could be separated purely in the wavelength domain (i.e., not decorrelating versus proxies for magnetic activity).  \citet{Collier-Cameron_2021} used a cross-correlation technique to measure solar RVs, and indeed cross-correlation has been widely used to measure precise RVs, and the technique has historical roots that go back half a century to \citet{Griffin_1967}.  Recently, line-by-line techniques are being increasingly explored \citep{dumusque_2018, Cretignier_2020a, Artigau_2022, Moulla_2022}, and have distinct advantages.  The effect of magnetic activity can vary qualitatively and quantitatively among different spectral lines, for example as a function of line depth \citep{Cretignier_2020a, Palumbo_2022}. Line-by-line techniques can thereby help to define and correct for varying manifestations of magnetic activity.

In this paper, we explore how to separate center-of-mass RVs from magnetic effects using line bisectors in the solar spectrum.  Measurement of accurate line-by-line bisectors requires very high spectral resolving power \citep{Lohner-Bottcher_2019}, and that in turn requires a high photon flux.  While solar measurements are not photon-starved, stellar observations of line bisectors could be challenging.  Nevertheless, the new generation of Extremely Large Telescopes \citep{Hook_2004} could enable bisector analyses of stellar spectra, if the appropriate high resolution spectrometers can be deployed.  To develop a bisector technique for EPRVs, we study bisectors in solar integrated light spectroscopy over a full magnetic cycle, as observed by the SOLIS project \citep{Keller_1998}.  We investigate how line bisectors are perturbed by solar magnetism, and we explore how to use bisectors to separate magnetic effects and center-of-mass RVs. 

\subsection{Organization of This Paper}

This paper is organized as follows.  Section~\ref{sec: solis} describes the data that we use from the SOLIS project \citep{Pevtsov_2014}, and compares it to solar integrated light spectra from other instruments.   We derive line bisectors from the SOLIS data, and we also measure the effect of telluric absorption \citep{Cunha_2014}, in Section~\ref{sec: bisectors}.   Section~\ref{sec: analysis} describes our analysis and results, including how we evaluate the shape of the bisectors (Section~\ref{sec: span}), how bisectors vary over the solar cycle (\ref{sec: variations}), and how those magnetic effects can in principle be separated from center-of-mass RVs (\ref{sec: separating}).  Section~\ref{sec: limitations} describes some limitations of our analysis, and Section~\ref{sec: future} discusses the potential for future work.  Section~\ref{sec: summary} summarizes our results.

\section{SOLIS Spectra}\label{sec: solis}

SOLIS is the Synoptic Optical Long-term Investigations of the Sun \citep{Keller_1998}, a project of the National Solar Observatory\footnote{https://nso.edu/telescopes/nisp/solis}.  SOLIS acquires magnetograms and narrow-band images in addition to spectra of the Sun in integrated light using the Integrated Sunlight Spectrometer (ISS, \citealp{Pevtsov_2014}).  ISS spectra are acquired in 10 segments of width varying from 0.5 to 1.6\,nm \citep{Bertello_2011}.  Some of those segments contain strong saturated lines having chromospheric components such as the Ca H\&K lines that are not suitable for our purpose.  Other segments are affected by prominent telluric absorption (e.g., near the sodium D1 line), making it impractical to measure reliable line bisectors.  Our analysis is based on measuring 62,325 bisectors of 21 lines in 5 segments.  The spectral lines and our results are summarized in Table~\ref{table: list}.  

The ISS spectra are not wavelength-calibrated using a laser comb or similar device (although one segment uses an Iodine cell).  Rather, the spectra are transformed in wavelength \citep{Bertello_2011} to agree with integrated light spectra taken with the McMath-Pierce Fourier Transform Spectrometer (FTS) that have very high signal-to-noise ratio and spectral resolution \citep{Livingston_2007}.  The absolute wavelength calibration of the FTS \citep{Reiners_2016} is not important for our analysis because we are analysing only the {\it shapes} of the lines in the ISS spectra, and a simulated RV signal that is injected (Section~\ref{sec: separating}).  Our analysis is thereby done in the heliocentric reference frame, as noted in Sections~\ref{sec: bisectors} and \ref{sec: separating}.  The ISS spectra have a spectral resolving power of 300,000, which permits excellent measurement of line bisectors \citep{Lohner-Bottcher_2019}.

\begin{figure}
\centering
\includegraphics[width=3in]{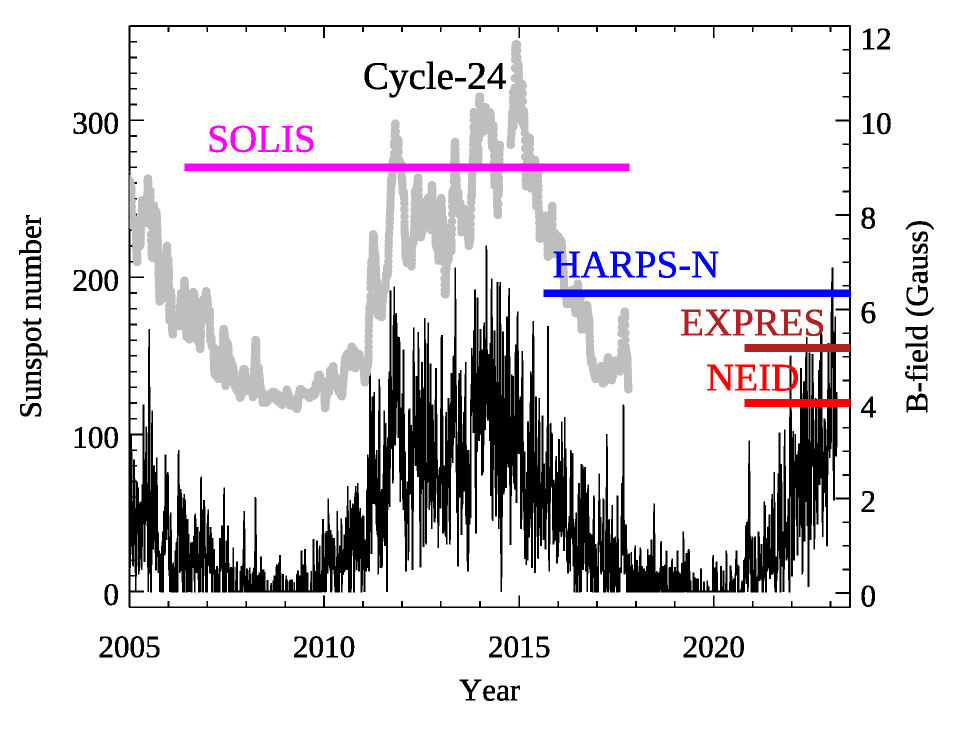}
\caption{Time span over which integrated light solar spectra were acquired by the SOLIS project, HARPS-North, EXPRES, and NEID.  
The left Y-axis is the number of sunspots, showing the extent of solar activity in Cycle-24, and now increasing in Cycle-25.  The points in grey are
the spatially-averaged magnetic flux on the solar disk, in Gauss as measured by the SOLIS Spectromagnetograph (read Y-axis on the right). \label{fig: spotfig} }
\end{figure}

Figure~\ref{fig: spotfig} shows the time span over which SOLIS/ISS spectra were acquired, compared to sunspot numbers as a proxy for solar activity.  The time span for other integrated light solar spectra (HARPS-N, etc) are also illustrated.  There are two primary advantages of the ISS spectra for studying how magnetic activity affects the apparent RV of the Sun in integrated light.  First, the ISS spectra cover almost all of Cycle-24, so the effect over a complete activity cycle can be delineated.  Second, the high spectral resolving power (300,000) of the ISS spectra means that line bisectors can be measured with high spectral purity.

\begin{figure}
\centering
\includegraphics[width=3in]{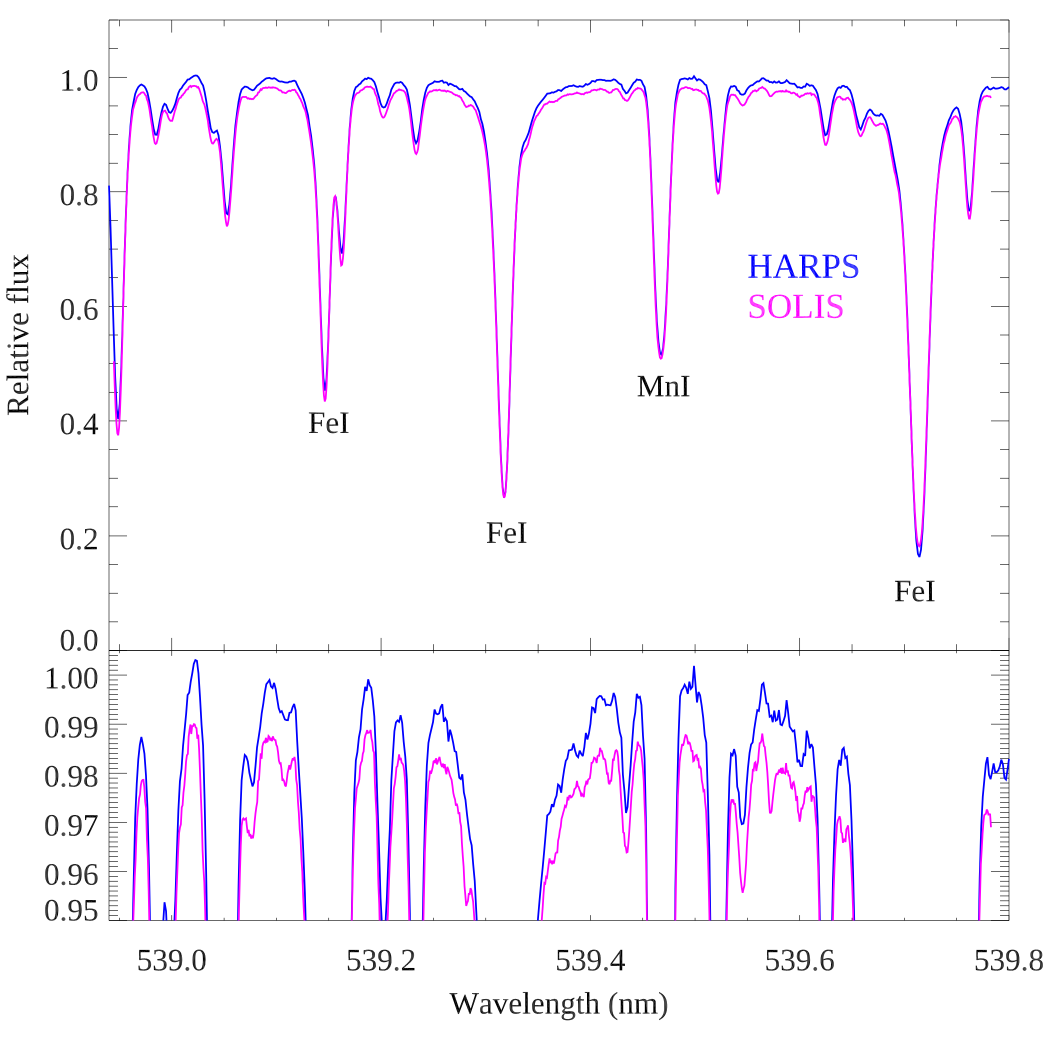}
\caption{Sample SOLIS spectrum covering the region near the MnI line at 539.5\,nm, compared to an average of HARPS-N spectra acquired on the same day.  Both spectra represent the full data for that single day of observing, randomly selected as August 17, 2015 (a high humidity season for the SOLIS data).  The lower panel expands the upper 5\% of the spectrum, so that the noise can be seen (compare fluctuations of the two spectra in regions near the continuum).  The SOLIS spectrum is offset downward by 1\% to facilitate the visual comparison.  \label{fig: sample} }
\end{figure}

Figure~\ref{fig: sample} shows an example of an ISS spectrum covering the neutral Manganese line near 539.5\,nm, compared to the average of HARPS-N solar spectra on the same day.  SOLIS typically acquires one spectrum per day in each band, whereas HARPS-N acquires multiple spectra.  Nevertheless, the signal-to-noise ratio per day is similar, as can be seen on Figure~\ref{fig: sample}.  Consequently, our analysis is based on spectra comparable in quality to the HARPS-N spectra that have been used in previous RV analyses of disk-integrated sunlight \citep{Collier-Cameron_2019, Milbourne_2019}. 


\section{Measuring Line Bisectors}\label{sec: bisectors}

 A line bisector is a simple concept: the X-values (wavelengths) in a bisector are a function of the relative flux level in the two flanks of the line profile.  A range of Y-values (fluxes) are adopted, and the X-value of the bisector is extracted from the data at each Y-value.  At a given Y-value, two X-positions are measured: where each flank of the line has exactly the given flux level.  The average of those two positions is the X-value of the bisector, for that Y-value of relative flux.  The first requirement for a bisector measurement is that the instrumental response must be removed, so that the flux level of the continuum reaches the astrophysical value.  That astrophysical continuum flux value is usually normalized to unity, as in Figure~\ref{fig: sample}.  With the continuum level rectified, measurement of a line bisector requires only interpolation in the measured flux versus wavelength.

 The SOLIS/ISS spectra that we use are wavelength-calibrated to coincide with a heliocentric reference frame (at rest relative to the Sun), and the instrumental response is removed.  That reduction process, and details of the ISS spectrometer and its operation were described by \citet{Bertello_2011}.  Given the nature and quality of the ISS data reduction, our measurement of line bisectors requires only accurate interpolation in the calibrated spectra.  To start that process, we calculate a Fourier transform of each spectrum, and zero-fill those transforms to a Nyquist frequency 18 times greater than the highest frequency defined by the sampling of the observed data.  To avoid ringing from edges, we extend the observed data by appending a mirror image of the spectrum at each end, and apodize that extended spectrum using a cosine-bell function.  Then, an inverse Fourier transform yields a version of the original spectrum re-sampled to a much denser wavelength grid than the original observations.  For a given flux level (Y-value), we then find the corresponding X-value (i.e., wavelength) in the red and blue flanks of each line, using quadratic interpolation applied to the dense grid (even the dense grid from the Fourier process will not correspond exactly to a given Y-value).  The dense grid from the Fourier process is very helpful for interpolation generally, and specifically to measure bisectors into the line core, where other interpolation methods can struggle to derive accurate values.  However, we also 'anchor' each bisector in the line core by fitting a parabola to data within $\pm 0.004$\,nm of the lowest point in the line, and we include the minimum of that parabola as the lowest point in the line bisector.
  
 We tested our Fourier+quadratic interpolation process by injecting a synthetic line defined using an analytic Gaussian.  Because that line is constructed using an analytic expression, we know the values in the bisector exactly.  Applying our Fourier interpolation method, we derive a bisector closely equal to the analytic values, the standard deviation of the difference being 1 cm/sec in equivalent velocity.  Hence, the numerical process that we use to measure line bisectors does not add significant numerical noise to the analysis.

Line bisectors can be affected by the presence of other absorption lines nearby in wavelength, because those lines can depress the flux in the flanks of the line whose bisector we are measuring.  Even seemingly isolated absorption lines can be affected to a small degree by this effect.  To minimize it, we follow our interpolation process with a fit and removal of a straight line to each bisector, i.e., we remove wavelength variations that are a linear function of flux level. Consequently, subsequent measurements of bisector properties (Section\ref{sec: span}) will be focused solely on changes in the {\it curvature} of the bisectors.
 
\begin{figure}
\centering
\includegraphics[width=3in]{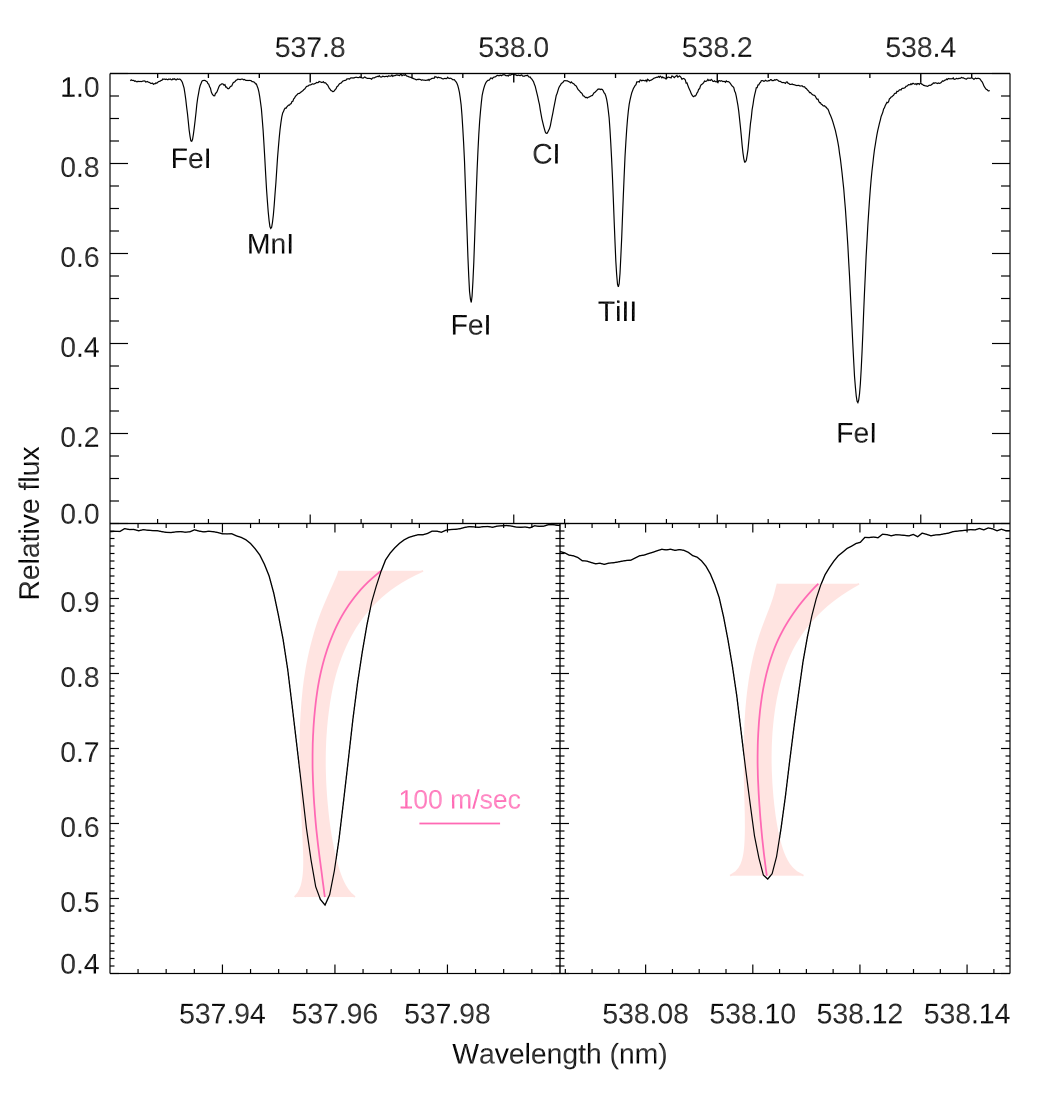}
\caption{Examples of bisectors that we measure using the SOLIS/ISS spectra.  This example is for the spectral segment centered on the high-excitation line of neutral carbon at 538.032\,nm.  The two bottom panels expand two lines (left: FeI 537.957\,nm; right: TiII 538.102\,nm), with the average bisector for each line overplotted (in hot pink) on an even more expanded scale (refer to the 100 meter/sec legend).  The bisectors plotted are average shapes over 3587 spectra.  The rose-shaded region surrounding each bisector is the $\pm3\sigma$ uncertainty envelope of a {\it single} bisector measurement, and note the velocity scale in meters/sec.  \label{fig: examples} }
\end{figure}

Average bisectors for two lines are illustrated in Figure~\ref{fig: examples}.   The list of lines or which we derive bisectors are given in Table~\ref{table: list}.  For each line, Table~\ref{table: list} includes the maximum and minimum flux levels of the bisector.

\subsection{Bisector Uncertainties in Wavelength} 

Figure~\ref{fig: examples} shows the $\pm\,3\sigma$ uncertainty range in wavelength of a typical single bisector at each flux level, as a shaded region.  We calculated the uncertainty range starting with the noise level in flux of each spectrum.  We measure the noise level from the point-to-point fluctuations in the spectrum in a near-continuum region of wavelength where line absorption is minimal.  Using that noise level in flux, we convert it to a per-point uncertainty at every point in the bisector, dividing by the observed slope $\partial{F}/\partial{\lambda}$, where $F$ is flux and $\lambda$ is wavelength.  We assume that uncertainties in the blue- and red-flanks of each line contribute independently (i.e., add quadratically) to the uncertainty in the wavelength of the bisector.  The uncertainties measured in this manner are used when fitting a polynomial to measure the span of the bisector (see Section~\ref{sec: span}), and when fitting to separate center-of-mass RV from activity effects (Section~\ref{sec: separating}).  However, these uncertainties are not useful when measuring temporal changes in the shape of each bisector, because intrinsic changes in shape greatly exceed the error levels of the bisector wavelengths (see Section~\ref{sec: span}).

\subsection{Telluric Contamination}\label{sec: telluric}  

Contamination by telluric water vapor absorption \citep{Cunha_2014, Bedell_2019, Wang_2022, Ivanova_2023} can in principle affect many of the solar absorption lines in the SOLIS/ISS bands.  To evaluate what lines are affected, we measure the strength of the time-variable telluric absorption and then seek correlations of the varying solar line properties with the varying telluric strength.  For the strength of telluric absorption, we choose the most appropriate water vapor line in each spectral segment and we fit a Gaussian profile to measure its depth in each spectrum.  By most appropriate line, we mean a line sufficiently deep to measure precisely (e.g., $\sim 10$\%), but unsaturated and with reasonably clean flanking continuum regions.  
 
\begin{table*}[h]

    \centering\caption {Lines in the SOLIS/ISS spectra for which we measure bisectors.  Line
identifications are from \cite{Kurucz_1995} or \citet{Moore_1966}. E is the lower state
energy in electron volts.  The median bisector span in meters/sec over all times is listed as $S$, and $\Delta{S}$ is the
difference in median bisector from quiet to active periods (see text), in the sense of quiet minus active. $F_{min}$ and
$F_{max}$ are the minimum and maximum flux levels relative to the local continuum used for each bisector.}
\begin{tabular}{lllrrlll}
$\lambda$\,(nm in air) &  Species & E & $S$ (m/s) & $\Delta{S}$ (m/s) & $F_{min}$  &  $F_{max}$  & Notes \\
\hline 
\hline
537.683 &  FeI &  4.29  &  $43.1$ & $3.8\pm1.0$  &  0.85 & 0.97 &  \\ 
537.764 &  MnI &  3.84 &   $167.2$ & $1.8\pm0.5$ &  0.65 & 0.90 &  \\
537.957 &  FeI &  3.69 &   $26.3$ & $1.1\pm0.7$  &  0.48  & 0.95 &  \\
538.032 &  CI  &  7.68 &   $49.5$ & $-0.6\pm2.3$ &  0.87  & 0.97 &  \\ 
538.102 &  TiII & 1.57 &   $24.4$ & $2.3\pm0.5$  &  0.51 & 0.93  &  \\
538.177 &  CoI  & 4.25 &   $86.4$ & $-12.3\pm3.3$ &  0.95 & 0.975 &  \\
538.227 &  CaI  & 4.44 &   $46.9$ & $1.0\pm1.3$ & 0.80 & 0.97  & (1)  \\ 
538.337 &  FeI  & 4.31 &   $-72.0$ & $-3.7\pm0.4$ &  0.23 & 0.75  &  \\
539.047 &  CoI  & 4.06 &   $-46.2$ & $-1.1\pm1.1$ &  0.75  & 0.88 & \\
539.146 &  FeI  & 4.15 &   $30.0$  & $1.1\pm0.7$  &  0.44  &  0.77 &  \\ 
539.233 &  NiI  & 4.16 &   $19.3$ & $4.5\pm1.4$  &  0.88 & 0.97 &   \\
539.318 &  FeI  & 3.24 &   $40.6$ & $5.0\pm0.7$  &  0.26  & 0.85 &  \\
539.467 &  MnI  & 0.00 &   $11.8$ & $4.5\pm1.3$  &  0.51  & 0.96  &  \\ 
539.522 &  FeI  & 4.44 &   $16.1$ & $2.4\pm0.9$  &   0.81 & 0.96 &  \\
539.623 &  TiII & 1.58 &   $39.9$ & $3.0\pm1.5$  &  0.89  & 0.96 & \\
539.706 &  TiI  & 1.88 &   $-94.8$ & $-0.5\pm0.8$ & 0.17  & 0.79 & blend  \\
539.713 &  FeI  & 0.91 &    &   &   &  &  blend  \\
539.762 &  FeI  & 3.63 &   $-36.1$ & $1.6\pm1.0$  &  0.76 & 0.91 &   \\
655.959 &  TiII & 2.05 &   $66.8$ & $-0.1\pm1.2$  &   0.75 & 0.85 &   \\
853.802 &  FeI  & 4.91 &   $100.3$ & $-7.9\pm2.8$ &  0.79 & 0.83 &    \\
854.214 &  CaII & 1.70 &   $-572.3$ & $-54.2\pm5.1$  &  0.20 & 0.41 &  \\
1082.709 &  SiI & 4.96 &   $-86.3$ &  $-19.9\pm2.5$  &  0.48  &  0.80  & \\ 
\hline
\end{tabular}

\begin{tablenotes}
\item{
 Note: (1) \citet{Kurucz_1995} list the wavelength as 538.233, and the 0.06\,nm discrepancy exceeds
the observational errors.  We adopt the identification due to the lack of other transitions. }
\end{tablenotes}
\label{table: list}
\end{table*}

\section{Analysis and Results}\label{sec: analysis}

\subsection{Span of the Bisectors and Uncertainties}\label{sec: span}

We define the span of a line bisector to be the distance in wavelength from the line core to the inflection point in the bisector, in the sense of core wavelength minus inflection wavelength.  For bisectors having a "C-like" shape \citep{Dravins_1981}, the inflection point is the bluest point in the bisector.  For bisectors having a "reversed-C" shape \citep{Gray_2010}, the inflection point is the reddest point in the bisector.  In both cases we determine the inflection point by fitting a polynomial to the bisector, and calculating the point where the derivative of that polynomial is zero. The derivative is $\partial{\lambda}/\partial{F}$, where $\lambda$ is wavelength and $F$ is flux relative to the continuum.  We use polynomials of the lowest degree that fits each bisector within its error envelope, constraining the degree of the polynomial to be the same for all bisectors of a given line.  We find that degrees $N=2$ (quadratic) to $N=4$ are needed, depending on the strength of the line and the complexity of its bisector.  After determining the bisector span for all measurements of a given line, we test for correlation versus the strength of telluric absorption by calculating the Pearson correlation coefficient, and we reject lines having a clear correlation.

\begin{figure*}
\centering
\includegraphics[width=6in]{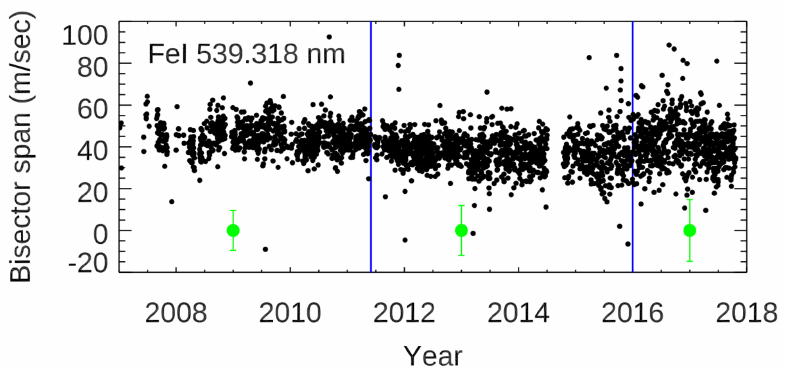}
\caption{Span of the bisector of the FeI line at 539.318\,nm, versus date.  The vertical blue lines mark the approximate times between which the Sun was most active in Cycle-24 (compare to Figure~\ref{fig: spotfig}).  In addition to the decrease in bisector span during the middle period of high solar actitvity, there are also fluctuations due to telluric effects (Section~\ref{sec: telluric} and Figure~\ref{fig: RV_telluric}). The green points with error bars are not real data, they illustrate the $\pm3\sigma$ envelope due to random variations per point, in the three regions. }
\label{fig: spantime}
\end{figure*}

Figure~\ref{fig: spantime} shows how the span of the bisector of the FeI line at 539.318\,nm varies with time over a solar cycle.  The median value of the span for this line is +41 meters/sec (Table~\ref{table: list}), indicating that the line core is at a greater wavelength than the inflection point, a classic C-shaped bisector.  The vertical blue lines on Figure~\ref{fig: spantime} mark the dates between which the Sun was most active in Cycle-24, and the bisector span was slightly reduced during that period, as we discuss below.  This Figure also shows some evidence for telluric effects on a yearly cycle due to the high humidity during the summer on Kitt Peak.  Nevertheless, over the long time scale of the ISS data, telluric contamination has an insignificant effect on the bisector span, as measured using the correlation coefficient.  (The same cannot be said regarding the RV state of this line, as discussed in Section~\ref{sec: separating}.) 

\begin{figure}
\centering
\includegraphics[width=3in]{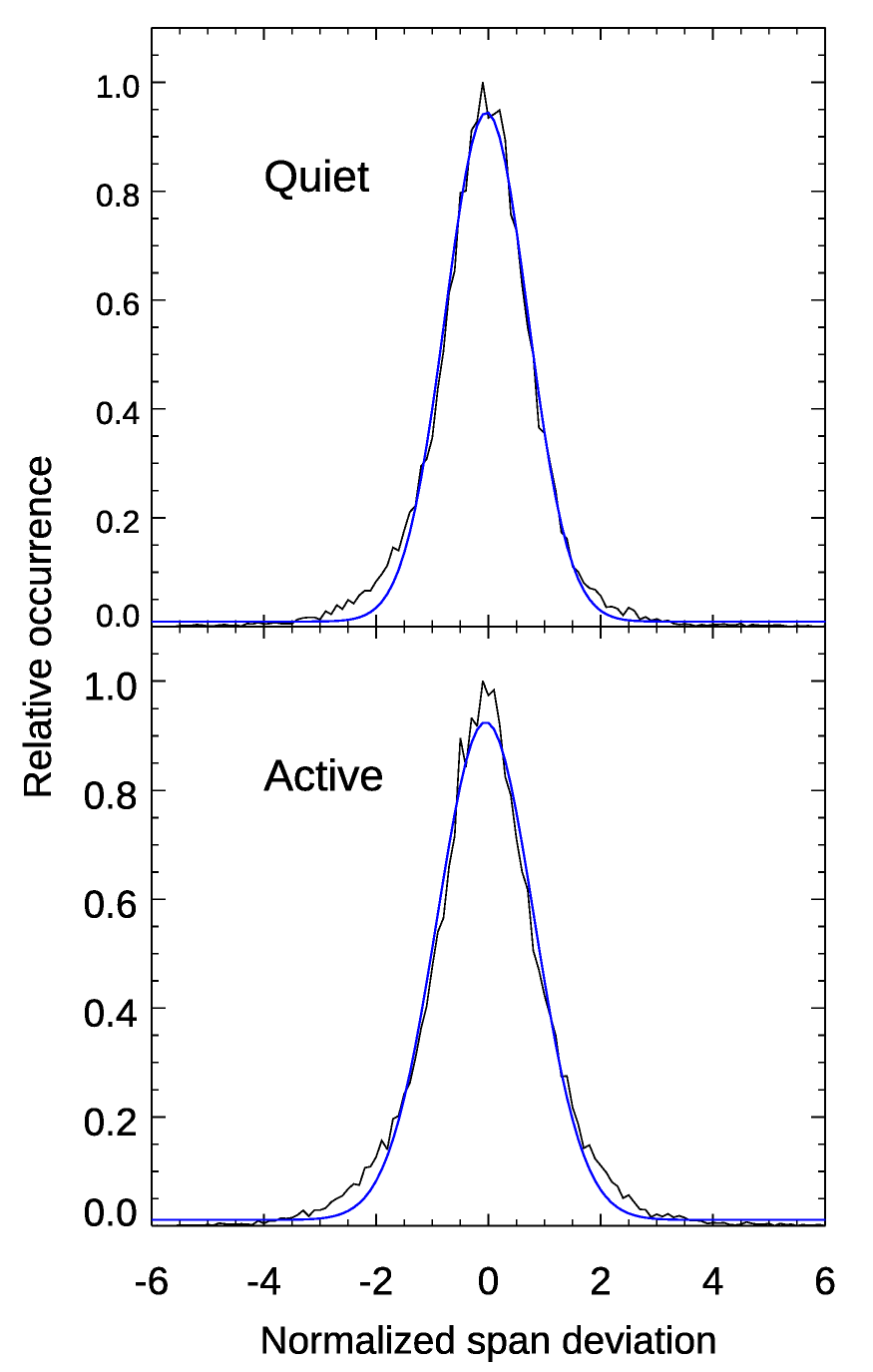}
\caption{Distribution of deviations in the bisector span, normalized by their observed standard deviation and divided into quiet-Sun and active-sun periods.  The measured distributions include all of the lines in Table~\ref{table: list}.  The blue curves are fitted Gaussian distributions. }
\label{fig: span_distrib}
\end{figure}

For all of the lines we study, the bisector span exhibits significant scatter as a function of time, exceeding the errors that we propagate from the signal-to-noise ratio in the spectra (Section~\ref{sec: bisectors}).  To determine to what degree bisector variations are related to photospheric magnetic activity, it is necessary to understand the statistical properties of this scatter.  For each line, we tabulated the deviations of the bisector span from their median value, and normalized those deviations by their standard deviation.  Normalizing the scatter in the bisector span by their standard deviations allows us to combine all of the lines into a single statistical distribution, and thereby define that distribution to high precision.  Figure~\ref{fig: span_distrib} shows the distribution of the span variations for all lines, divided into quiet-Sun and active-Sun periods.  We fit Gaussian normal error functions to the observed distributions, and we retrieve standard deviations close to unity (as expected, because the scatter was normalized).  Based on the Figure~\ref{fig: span_distrib} we find that the deviations of bisector span from their median values follow Gaussian error distributions to a good approximation.  However, that does not test for possible autocorrelation in the span values, and autocorrelation will strongly affect uncertainties when measurements are binned (as in quiet-Sun versus active-Sun).  Indeed, Figure~\ref{fig: spantime} suggests autocorrelation in time, and that would be reasonable given the nature of magnetic generation and decay on the Sun, as well as solar rotation. 

The most important dimension of possible autocorrelation in solar bisector spans is the magnetic dimension, i.e. how deviations in bisector span behave when they are binned according to the average magnetic flux on the solar disk.  For a given range of the magnetic flux, we expect that the scatter of the average bisector span should decrease as the square-root of the number of measurements that are averaged, the Allan deviation relation \citep{Allan_1966}.  For each line, we grouped the deviations in bisector span into two groups, quiet and active (see below).  We measured the slope of the relation between the standard deviation of average span deviations and the log of the number of measurements that are averaged.  Considering that slope over all 21 lines that we study, and over both the quiet and active groups, we find an average slope of -0.43 versus -0.5 for a perfect Allan deviation relation.  We use the slope of -0.43 when calculating error bars on the average difference between bisector spans at quiet and active epochs (Table~\ref{table: list}).

In addition to magnetic effects on the bisector span, we consider the possible effect of the 5-minute oscillation.  The ISS spectra use short exposure times (typically, 40 seconds) and that will not average over the 5-minute oscillation.  However, the amplitude of the 5-minute oscillation is significantly less than 1 meter/sec in whole disk spectra \citep{Strassmeier_2018}, approximately two orders of magnitude less than in disk resolved observations.  Even if the the 5-minute oscillation altered the bisector span by as much as the oscillation's velocity amplitude (which it does not, \citealp{gomez_1987}), it would not be a significant factor adding to variations in the line bisector spans that we measure. 

\subsection{Bisector Variations over the Solar Cycle}\label{sec: variations}

There is substantial previous work on the properties of line bisectors in the solar spectrum.  Investigators have exploited the resolved solar disk to measure how line bisectors vary in magnetic versus quiet regions, and how they vary from disk center to limb \citep{Cavallini_1982, Marmolino_1987, Neckel_1990,  Uitenbroek_2006, Lohner-Bottcher_2019}.  Investigation of bisectors properties in disk-integrated light has also been pursued \citep{Livingston_1983, Livingston_1984, Pietarila_2011, Giampapa_2015, Osipov_2017, Sheminova_2020}, but the behavior of multiple line bisectors over an 11-year solar cycle has yet to be adequately defined.  \citet{Livingston_2007} studied strengths of photospheric and chromospheric lines over a 30 year interval, but they focused on line strengths, not bisectors.  \citet{Giampapa_2015} studied bisectors of FeI 539.318 and MnI 539.523 in the ISS data from 2006 to 2014, and they found changes in the bisector of the MnI line, but not the FeI line.  Our work expands on previous studies of the Sun-as-a-star by measuring bisectors of 21 lines in the ISS data over an 11-year baseline covering Cycle-24.  We believe that our method of measuring bisectors, including Fourier interpolation to a very dense wavelength grid (Section~\ref{sec: bisectors}), produces accurate and sensitive bisectors.  Among other conclusions, we find a very clear variation in the bisector of the FeI 539.318 line, as well as other lines, as a function of solar activity (see below).

In order to define variations in line bisectors as a function of solar activity, we divide our measured bisectors for each line into two groups: quiet and active.  We use the full-disk average of the total magnetic flux, $\langle{\mid{B}\mid}\rangle$, measured by the SOLIS Vector Spectromagnetograph \citep{Henney_2009}, and we interpolate those magnetic flux values to the times of each bisector measurement.  When we examined bisector span as a function of magnetic flux, we found a natural division at a flux value of 6 Gauss, and this division is also consistent with Figure~\ref{fig: spotfig}.  We constructed a histogram of magnetic flux values, and that also suggested that 6 Gauss was a natural boundary between quiet and active phases.  Sorting the bisector spans by magnetic flux value is more physically motivated than a division by time (as in Figure~\ref{fig: spantime}).  Thus, our quiet phase represents measurements when $\langle{\mid{B}\mid}\rangle$ is less than 6 Gauss, and active phases are when $\langle{\mid{B}\mid}\rangle$ is greater than or equal to 6 Gauss.  For each line, we calculated the median bisector span independent of magnetic flux, and also the median span values in the quiet and active phases.  Using the Allan deviation slope of $-0.43$ (see above), we calculated the uncertainty of the median span value in each phase, and the uncertainty of the difference by combining the uncertainties for each phase quadratically.  The median bisector span over all data, and difference in span between quiet and active phases, are given for each line in Table~\ref{table: list}.

Table~\ref{table: list} shows that 15 of the 21 lines we study have C-shaped line bisectors (positive median span values).   Seven of those 15 lines have statistically significant ($>3\sigma$) changes in bisector span in active versus quiet phases.  All but one of those seven lines have changes in bisector span that are positive (quiet span greater than active span).  (The exception is the very weak line of CoI at 538.177\,nm.)  Six of the lines we measure have reversed-C bisectors.  Four of those six have statistically significant changes in active versus quiet phases, and in all four of those lines the change ($\Delta{S}$) is negative, implying that the active bisector is closer to symmetric than is the quiet bisector.  

We repeated our analysis using 5- and 7 Gauss as the dividing point between quiet and active phases.  Although the results for individual lines vary, depending on the exact value of the dividing point, our overall results are very similar when using 5- or 7 Gauss as the boundary between quiet and active phases.  Considering all of the bisectors we measure, our results overwhelmingly indicate that the asymmetries of line profiles in the spectrum of the Sun-as-a-star (both C-shaped and reversed-C-shaped bisectors), are reduced during periods of enhanced solar activity.  

\begin{figure}
\centering
\includegraphics[width=3in]{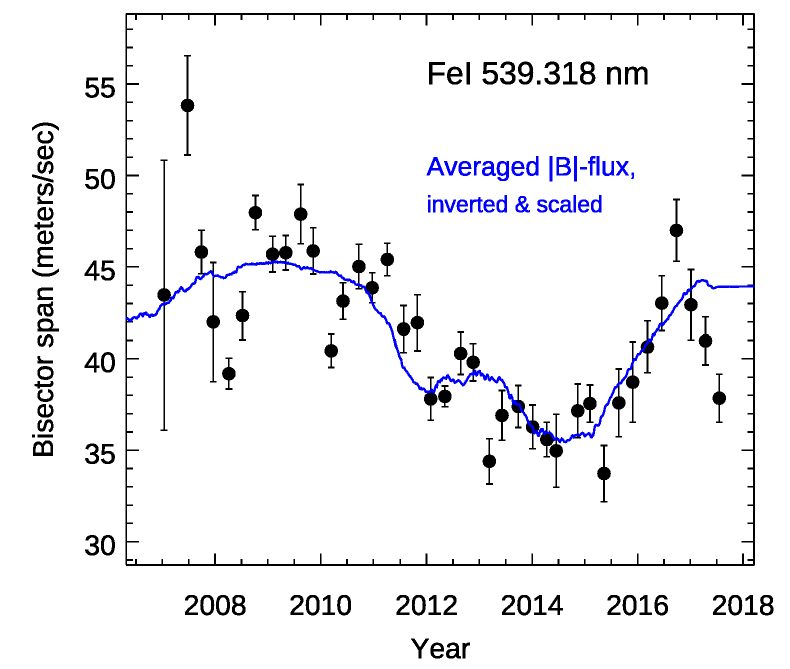}
\caption{Bisector span (in meters/sec) for the FeI line at 539.318\,nm versus date.  The blue line is the disk-average of the total magnetic flux, plotted on an inverted scale (greater flux being downward) and matched to the amplitude of the bisector variations.  The points are 100-day averages of the bisector span, and their scatter exceeds the error bars due to temporal variations on times less than the 11-year magnetic cycle.  }
\label{fig: 539bisector}
\end{figure}

\begin{figure}
\centering
\includegraphics[width=3in]{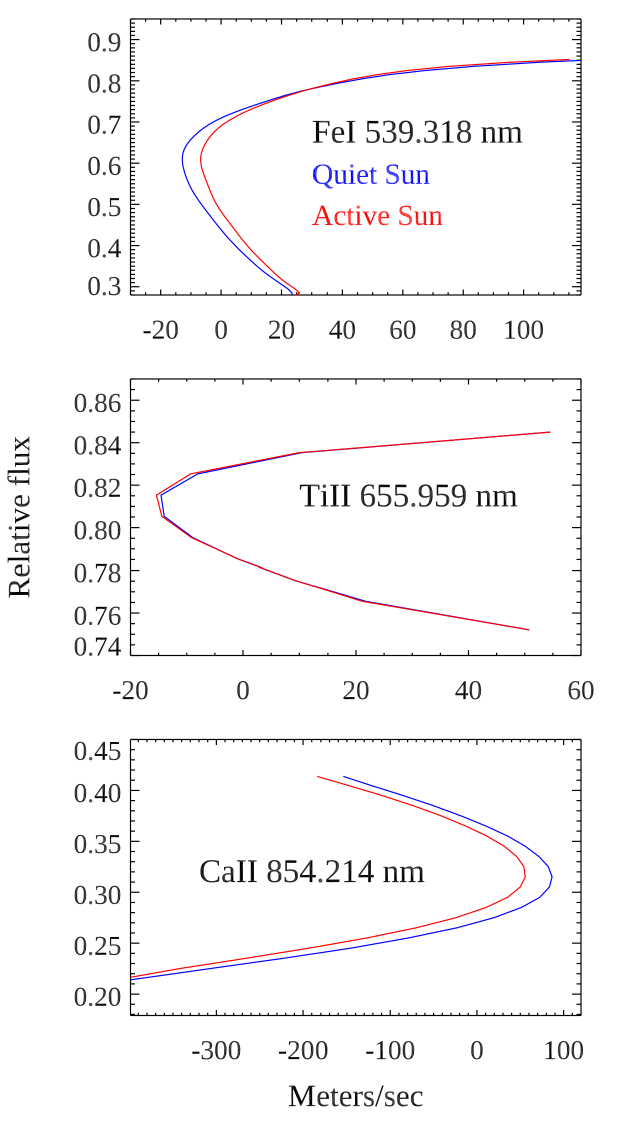}
\caption{Average line bisectors (in meters/sec) for the same three lines whose span is plotted versus $\langle{\mid{B}\mid}\rangle$ in Figure~\ref{fig: span_vs_B}.  The active and quiet bisectors for TiII 655.959\,nm on the middle panel are almost identical and overlapping at this scale.  Only the spans of the bisectors are meaningful here; the zero point on the X-axis is affected by imprecision in the rest wavelengths of the lines.  The $\pm1\sigma$ uncertainties on the difference in bisector span (quiet minus active) are listed in Table~\ref{table: list}, and are: 0.7, 1.2, and 5.1 meters/sec from the top to bottom panels, respectively.  Lines in the top and bottom panels exhibit a statistically significant change in bisector span from quiet to active phases. }
\label{fig: 3lines}
\end{figure}

\begin{figure*}[h]
\centering
\includegraphics[height=6in]{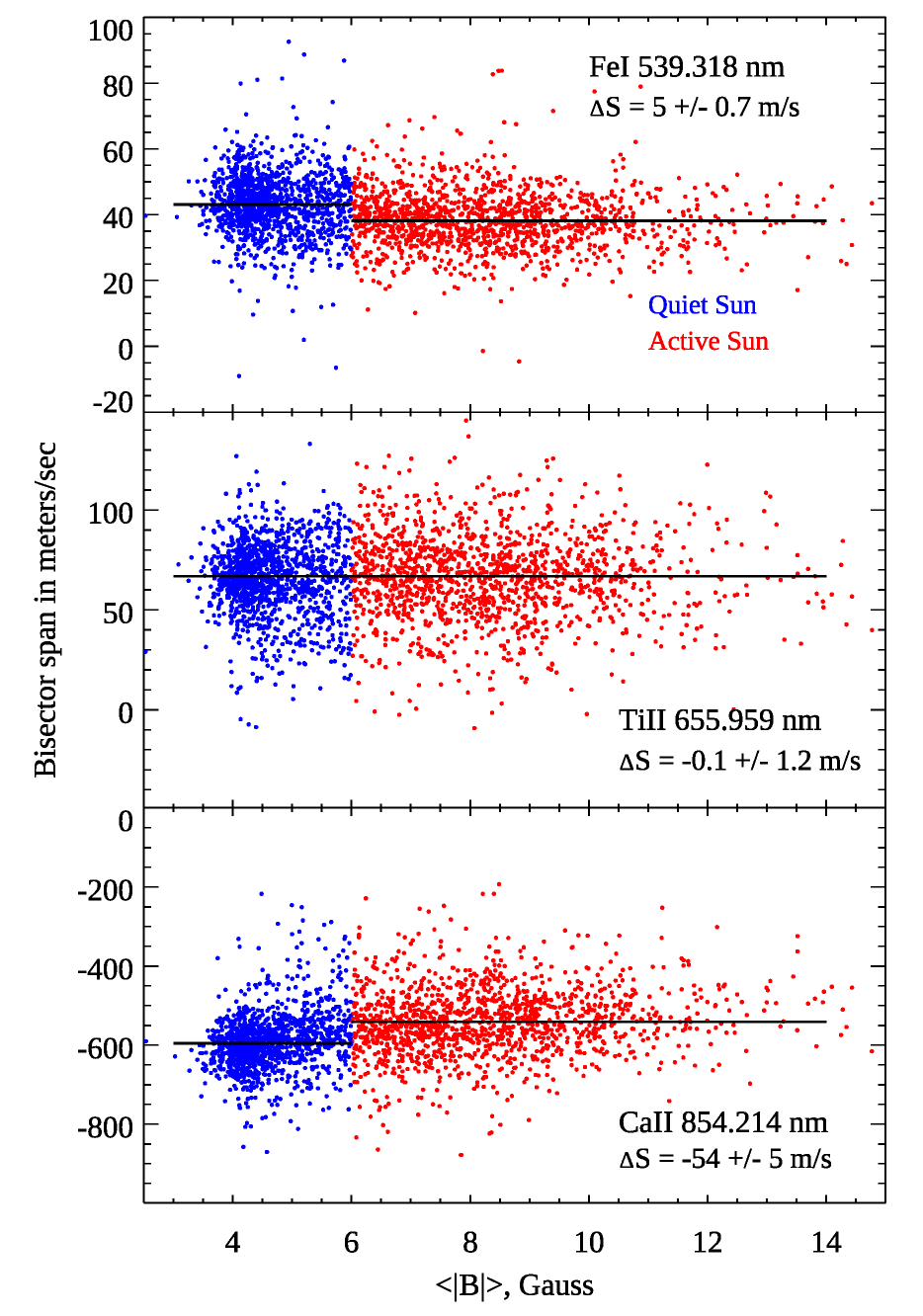}
\caption{Bisector span versus the disk-averaged total magnetic flux in Gauss, for three representative lines.  The top panel illustrates the line having the clearest change in C-shaped bisector (positive span).  The bottom panel shows the line with the clearest change in reversed-C bisector (negative span).  The middle panel shows a line with no change in bisector span to high precision.  For all lines, we divide the bisector measurements into quiet and active phases at a boundary of 6 Gauss.  Thus, quiet bisectors (blue points) have $\langle{\mid{B}\mid}\rangle < 6$\,Gauss, and active bisectors (red points) have $\langle{\mid{B}\mid}\rangle \geqq 6$\,Gauss. The horizontal lines indicate the median values in each phase. }
\label{fig: span_vs_B}
\end{figure*}

The FeI line at 539.318\,nm shows the clearest example of changes in bisector span that are related to $\langle{\mid{B}\mid}\rangle$.  For this line, the variations in bisector span are sufficiently clear that they can be seen as an explicit function of time.  Figure~\ref{fig: 539bisector} shows the bisector span of FeI 539.318 versus time, with $\langle{\mid{B}\mid}\rangle$ scaled and overplotted; the similarity of the temporal variation of bisector span and magnetic flux is obvious in this Figure.  Another way to visualize how bisector span varies with solar activity is to plot the span values versus $\langle{\mid{B}\mid}\rangle$.  Figure~\ref{fig: span_vs_B} shows this plot for three representative lines: one whose positive span is prominently reduced in the active phase (top panel, FeI 593.318\,nm), one with a prominent reversed-C bisector whose span also moves closer to zero in the active phase (CaII 854.214\,nm, bottom panel), and one whose bisector is unchanged from quiet to active phases (TiII, 655.959\,nm, middle panel).  Figure~\ref{fig: 3lines} shows the measured bisectors for these three lines, during the quiet and active phases.  The TiII 655.959\,nm line being unchanged as a function of $\langle{\mid{B}\mid}\rangle$ illustrates an important point: although our results indicate that line asymmetries are reduced during active periods, that does not imply that {\it all} lines participate. A corollary is that EPRV studies will ultimately have to consider stellar spectra on a line-by-line basis in order to make the most effective corrections for stellar activity.

Finally, we comment on the properties of reversed-C-shape bisectors in the ISS data, i.e. lines having negative bisector span values.  The most prominent of the reversed-C bisectors is for CaII 854.214\,nm.  This line is well known to be sensitive to the chromosphere \citep{Pietarila_2013}, and its bisector was measured over a solar cycle by \citet{Pietarila_2011}.  Our results for the large negative span of the bisector, becoming less negative during active periods, are in good quantitative agreement with the results of \citet{Pietarila_2011}.  In addition, we find reversed-C bisectors in several other lines, such as SiI 1082.709\,nm and FeI 538.337\,nm.  Although reversed-C bisectors are umcommon in the solar spectrum \citep{Dravins_1981}, they are more common in the ISS spectra probably because the ISS spectral segments were preferentially selected to cover strong lines such as CaII 854.214\,nm.  It is significant that these bisector spans become less negative during active periods. The physical cause of reversed-C bisectors has been puzzling \citep{Gray_2010}, but there is good reason to believe that the cores of these strong lines are shaped by acoustic shock waves that propagate upward from the photosphere \citep{Uitenbroek_2006, Pietarila_2013}.  The bisector span of CaII 854.214\,nm has long been known to become less negative in active regions \citep{Pietarila_2011, Pietarila_2013}; our results indicate that the same is true for other strong lines with reversed-C bisectors.

\subsection{Separating Magnetic and RV Signals}\label{sec: separating}

A principal motivation for this work is to investigate whether extremely precise RVs for solar-type stars can be obtained using line bisectors, including the separation of center-of-mass RVs from line profile changes induced by photospheric magnetism.   In principle, a line bisector is a direct probe of the center-of-mass RV.  In the ideal case of a symmetric absorption line, the line bisector reduces to a vertical line at exactly the RV-shifted wavelength of the line, and immediately yields the center-of-mass RV.  In the realistic case where the line profile is perturbed by photospheric convection (modulated by magnetism), it is less obvious that RV changes can be separated from changes in the line profile that affect the bisector.  Perhaps the changes in line bisectors can be sufficiently large and complex to render it impossible to distinguish variations in their center-of-mass wavelength?  Since \citet{Collier-Cameron_2021} were able to accomplish that separation using cross-correlation RVs, we expect that it should be possible using a line-by-line approach such as bisectors.  Separation using techniques other than cross-correlation is yet to be demonstrated, but we do so here.

We injected a synthetic RV signal into the SOLIS spectra.  That signal is a sinusoidal RV variation having a period of exactly one year, and an amplitude of 9 cm/sec, the same as the Doppler reflex of the Earth's orbit.  For each line, we attempt to recover that signal in the presence of much larger variations in the line bisectors.  After adding the velocity signal, we decompose the collection of bisectors for each line using a principal component analysis.  We then fit each individual bisector using those principal components via multi-variate linear regression, retrieving also a constant offset in wavelength that is intended to recover the injected RV signal. The linear regression uses the uncertainty envelope of each bisector (e.g., Figure~\ref{fig: examples}) when converging to the best fit.  The number of principal components that we used varied from line to line.   For each line, we begin by fitting 3 components to every ($\sim$daily) bisector, and we calculate the scatter (standard deviation) of the retrieved offsets in wavelength.  As we increase the number of components used as basis vectors in the linear regressions, the scatter of the retrieved offsets decreases and approaches a minimum value.  We adopt the least number of components as basis vectors to reach the minimum scatter in the results, avoiding over-fitting.  That typically requires between 3 and 12 components, depending on the strength of the line and the complexity of the bisector shapes. The standard deviation of the retrieved offsets is always much larger than the amplitude of the injected RV signal. Nevertheless, we hypothesize that the injected signal can be recovered by averaging over multiple lines, and over the long baseline in time and range of solar activity that are sampled by the SOLIS data.   

\begin{figure*}[h]
\centering
\includegraphics[width=5in]{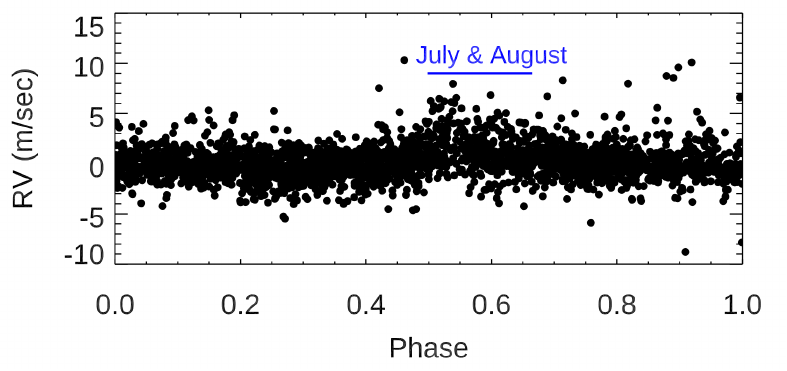}
\caption{Retrieval of the center-of-mass RV versus orbital phase of the injected signal, for the single FeI line at 539.318\,nm.  This shows RV results extracted from 2672 line bisectors measured during 10.8 years.  The standard deviation of the points plotted here is 1.7 meters/sec, so we do not expect to retrieve the injected signal without averaging over bins in phase, and also over multiple lines.  Because the injected signal has a one year period (with zero RV on January 1), the orbital phase plotted here is the same as the terrestrial yearly calendar.  The increase in humidity over Kitt Peak in the summer months affects the RV retrieved from this line, noticeable as increased scatter and a bias toward more positive (redder) RV values during July and August (see text for discussion). }
\label{fig: RV_telluric}
\end{figure*}

Figure~\ref{fig: RV_telluric} shows the retrieved RV results for the FeI line at 539.318\,nm.  This line provided the clearest example of a modulation of the span of the line bisector by the magnetic cycle (Figures~\ref{fig: 539bisector}, \ref{fig: 3lines} and \ref{fig: span_vs_B}), but RV results for this line are affected by micro-telluric contamination.  \citet{Moore_1966} lists a weak water vapor line at 539.367\,nm, sufficiently close to the solar line that it pulls the RV to more positive (redder) values during times of high terrestrial humidity.  Kitt Peak is well known to experience greater humidity in the summer months.  Because our injected RV signal has a 1-year period, starting from zero on January 1, the orbital phase of our hypothetical planet coincides with the terrestrial calendar.  Figure~\ref{fig: RV_telluric} shows this effect clearly as greater scatter occurs to positive RV during July and August, contaminating that range of phase in the signal of the hypothetical planet.  Accordingly, we drop this line from our RV analysis, but we deem the results for bisector span to be reliable.  As noted in Section~\ref{sec: span}, the long duration of the magnetic cycle compared to the yearly telluric variation results in negligible correlation of span values with the strength of telluric water vapor absorption.  We also drop CoI 538.177\,nm from the RV analysis because that line is too shallow for sensible regression fits to the bisector (even if we kept it, it's too weak to affect the RV results). 

\begin{figure}[h]
\centering
\includegraphics[width=3in]{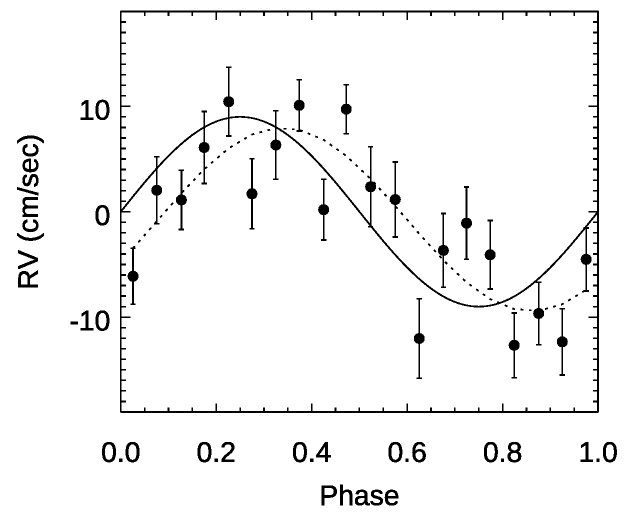}
\caption{Recovery of the injected RV signal with 9 cm/sec amplitude, by averaging over bisectors of 19 lines, and binning to a resolution of 0.05 in orbital phase. The solid curve is the injected signal, and the dashed line is the best-fit to the data and is consistent with the injected signal to within the errors (see text).}
\label{fig: veloc_result}
\end{figure}

We do not expect to detect the injected RV signal using single lines, because even the best lines have a point-to-point RV scatter that is much larger than the injected signal amplitude. However, we average over the thousands of bisectors for 19 lines (weighting each line by the inverse of its variance in RV), and we average over bins of 0.05 in phase. That allows us to recover the injected signal, as illustrated in Figure~\ref{fig: veloc_result}).  To further verify the recovery of the injected signal, we fit a sinusoidal RV curve (i.e., circular orbit) to the averaged and binned data, and the best-fit curve is shown as a dashed line on Figure~\ref{fig: veloc_result}.  We fit for the period, amplitude and a phase offset in the RV curve, and we derive uncertainties in those parameters using a bootstrap procedure.  We make $10^4$ realizations of the binned RV data, by adding random deviations from a Gaussian distribution to the best-fit curve, normalized to the observed error bars.  Fitting to those realizations of synthetic data, we calculate the standard deviation of the retrieved parameters.  On this basis we find best-fit period, amplitude, and phase offsets to be $1.04\pm0.10$\,years, $8.7\pm1.0$\,cm/sec, and $0.079\pm0.034$ offset in phase.  These results indicate that the injected signal is recovered to solid statistical significance and the recovered parameters agree with the injected curve to within the errors.  In principle, this result could depend more heavily on lines where the activity-related change in bisector span is minimal.  However, by inspecting the weights of the lines in the RV solution, we verified that the result in Figure~\ref{fig: veloc_result} depends approximately equally on lines wherein we detect a significant change in bisector span over the solar cycle, as well as lines whose bisector spans do not change within our errors (Table~\ref{table: list}).  We conclude that the small Doppler reflex of an Earth-like planet orbiting a sun-like star can be separated from activity-induced changes in line profiles using line bisectors.  That conclusion comes with some caveats, due to limitations in our analysis, as we now discuss.

\section{Limitations of Our Analysis}\label{sec: limitations}

The RV community has appropriately given great attention to the issue of accurate wavelength calibration \citep{Coffinet_2019, Zhao_2021, Gibson_2022, Kanodia_2023}.  Whereas, in our analysis we are insensitive to wavelength calibration because the SOLIS/ISS spectra are already shifted to a common reference frame (heliocentric).  Nevertheless, our analysis has other limitations.  First, we injected a synthetic signal instead of recovering a real signal.  It has been our general experience that synthetic signals are often easier to recover than are real signals.  (For example, real signals can be intertwined with unanticipated instrumental systematic effects.) Second, the continuum level is already determined, and the spectra shifted to the heliocentric frame, before we inject the synthetic signal.  Hence the signal cannot be attenuated by some critical first stages of the data analysis (e.g., continuum placement, \citealp{Cretignier_2020b}).   A real RV signal would be present in the data prior to all aspects of our analysis, and how we treat the data might then degrade the results.  In the case of stellar observations that are the ultimate goal, the RV changes due to the Earth's 30 km/sec motion could make especially large perturbations (e.g., stellar wavelength shifts relative to microtelluric lines) that may be difficult to correct.  Nevertheless, our analysis establishes that {\it activity-induced changes in the shapes of line bisectors are not in themselves} a fatal barrier to the RV detection of Earth-like planets orbiting Sun-like stars.   

\section{Future Work}\label{sec: future}

We are beginning a program to monitor line-by-line RVs using line bisectors in the spectrum of the Sun-as-a-star.  We plan to utilize solar spectra from NEID \citep{Lin_2022}, EXPRES \citep{Llama_2022} and HARPS-N \citep{Dumusque_2021}.  Because the Earth-Sun RV varies by $\sim\pm300$\,meters/sec over a year, we have the opportunity to extract that real signal to maximum accuracy, and thereby to remove the limitations mentioned in Section~\ref{sec: limitations}.  We will also identify the most stable lines as well as the lines most responsive to magnetic activity, and thereby develop line-by-line proxies that can be applied to stellar RV analyses.  
  
\begin{figure}[h]
\centering
\includegraphics[width=3in]{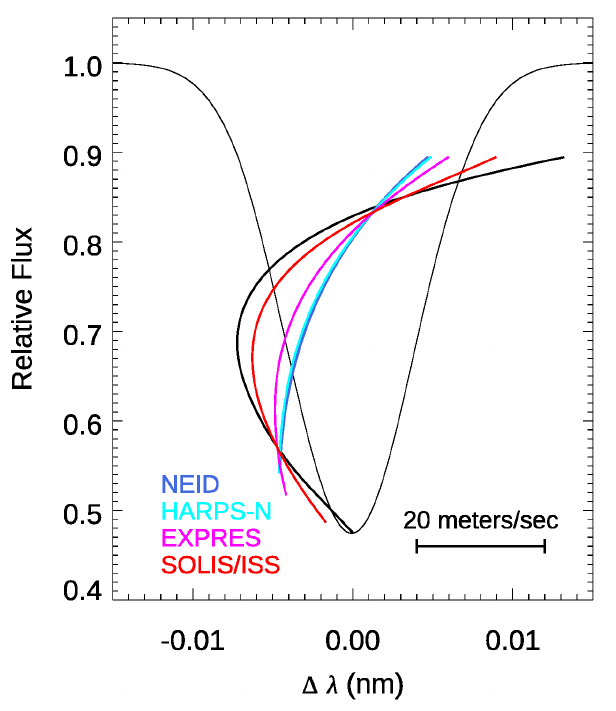}
\caption{Effect of spectral resolving power on line bisectors.  This shows a synthetic absorption line with a bisector whose shape agrees with solar measurements at very high resolving power. The line bisectors are plotted on a greatly expanded wavelength scale (see 20 meter/sec legend).  The black line is the fully resolved bisector, i.e. at infinite resolving power. After convolving to the spectral resolution of the NEID, HARPS-N, EXPRES, and SOLIS/ISS spectrometers, the span of the bisectors (colored lines) are reduced from the fully resolved case, but are still sufficiently prominent to serve as probes of solar activity and allow separation of center-of-mass RVs from activity-related RV variations.}
\label{fig: bisector_compare}
\end{figure}

Because our approach relies on line bisectors at high spectral resolution, we have investigated to what degree the spectral resolving powers of NEID, EXPRES, and HARPS-N will enable bisector measurements to high accuracy.  The SOLIS/ISS spectra have a resolving power of 300,000, but the solar feeds of NEID, EXPRES, and HARPS-N have lower resolving power by factors of two to three, and we explored to what degree the lower resolving power will affect bisector measurements.  We made a synthetic line using an analytic (Gaussian) profile that is closely matched to the width of real solar lines.  We distorted that profile by shifting the wavelengths at each flux level so that their centroid agreed with a typical (fully resolved) solar bisector.  We then convolved that asymmetric line profile to the resolution of NEID, EXPRES, and HARPS-N, and re-measured the line bisector.  The results are shown in Figure~\ref{fig: bisector_compare}, and it is apparent that the bisector spans measured by these stellar spectrometers will have reduced amplitudes relative to the spans measured at infinite resolving power.  Nevertheless, we believe that the stellar spectrometers have sufficient sensitivity to bisector shapes to permit our analysis technique to be successful.  

Using our line bisector approach, EXPRES is especially important due to its spectral resolving power (140,000).  Since late 2019, EXPRES has been acquiring spectra of the Sun-as-a-star almost every clear day at its location adjacent to the Lowell Discovery Telescope.  Each day's observations consist of a series of 200-second exposures, each achieving a signal-to-noise ratio of approximately 500.  Because the observations are nearly continuous while the Sun's altitude is above 15-degrees, the RV effect of differential extinction across the solar disk can be defined to high precision.  Currently, approximately 40,000 solar spectra from EXPRES await analysis of their line bisectors. 

\section{Summary}\label{sec: summary}

We have investigated whether spectral line bisectors can be used to separate center-of-mass RVs from false RVs caused by the modulation of photospheric convection by magnetic fields. We begin by measuring the properties of line bisectors in spectra of the Sun-as-a-star, acquired at high spectral resolving power (300,000) by the Integrated Sunlight Spectrometer (ISS) in the Synoptic Optical Long-term Investigation of the Sun (SOLIS, Section~\ref{sec: solis}). Implementing a Fourier interpolation method, we measure 62,235 accurate line bisectors for 21 lines observed with the ISS over an 11-year period covering the activity maximum of solar Cycle-24 (Section~\ref{sec: bisectors}). Telluric water vapor absorption is present in the ISS spectra to varying degrees. Measuring the depth of telluric lines, we seek correlations between bisector properties and the strength of telluric absorption, and thereby restrict our analysis to solar lines that are mostly free of telluric contamination (Sections~\ref{sec: telluric} and \ref{sec: span}).

We fit polynomials to the bisectors so as to measure their shapes without being mislead by outlying points.  We define the span of a line bisector to be the wavelength difference between the line core and the inflection point of the bisector.   A positive span corresponds to a C-shaped bisector, and 15 of the 21 lines we study have classic C-shaped line bisectors.   Disk-average magnetic fluxes were observed by the SOLIS Vector SpectroMagnetograph, approximately contemporaneous with the ISS spectra, and we correlate those fluxes with changes in bisector span (Section~\ref{sec: variations}).  We divide the bisectors into quiet and active phases, based on the disk-averaged magnetic flux, using 6 Gauss as the dividing point. Seven of the C-shaped bisectors have statistically significant ($>3\sigma$) changes in bisector span in active versus quiet phases.  Except for one weak line, the C-shaped bisectors exhibit changes in bisector span that are positive (quiet span greater than active span).  We also find that six of the lines we measure have reversed-C bisectors.  Four of those six exhibit statistically significant changes in active versus quiet phases, and in all four cases the active bisector is closer to symmetric than is the quiet bisector.  The reversed-C bisectors we measure include the chromospherically-sensitive line of CaII at 854.214\,nm, and our results are consistent with previous bisector studies of this line.  Motions and temperature fluctuations in the solar atmosphere cause spectral line asymmetries, and inhibition of those fluctuations by magnetic fields is expected to reduce those asymmetries.  Our results overwhelmingly confirm that the asymmetries of line profiles in the spectrum of the Sun-as-a-star (both C-shaped and reversed-C-shaped bisectors), are reduced during periods of increased solar activity.  

Our principal motivation is to determine whether changes in line shape as measured using bisectors can frustrate RV detection of Earth-like planets orbiting Sun-like stars, or whether changes in line shape can be separated from center-of-mass RVs.   We injected a synthetic RV signal into the bisectors.  The amplitude (9\,cm/sec) and period (1 year) of that signal were chosen to mimic the circular orbit of the Earth orbiting the Sun.  For each line, we apply a principal component analysis to its bisectors, and then we use the most significant of those components as basis vectors in a multivariate linear regression.  Those regression fits to each individual bisector include a constant term that is intended to capture the synthetic center-of-mass RV signal. Although we cannot recover such a small RV signal using any single line, an error-weighted average of 19 lines, binned in orbital phase, recovers the injected signal to high significance (Section~\ref{sec: separating}).  Although our work has limitations (Section~\ref{sec: limitations}), we conclude that activity-induced changes in the shapes of line bisectors are not in themselves a fatal barrier to the RV detection of Earth-like planets orbiting Sun-like stars.

\acknowledgements

This work utilizes SOLIS data obtained by the Integrated Synoptic Program, managed by the National Solar Observatory (NSO), which is operated by the Association of Universities for Research in Astronomy, AURA Inc. under a cooperative agreement with the National Science Foundation.   We thank Dr. Luca Bertello of NSO for guidance and assistance with the SOLIS spectra, and we thank an anonymous referee for comments that significantly improved this paper.  We gratefully acknowledge support by NASA's Extreme Precision Radial Velocity Foundation Science 2020 program via grant 80NSSC21K2007 to the University of Maryland.


\clearpage

\bibliography{references.bib}
\end{document}